\documentclass{LT23auth}
\usepackage{graphicx,amsmath,amssymb}

\begin{document}

\begin{frontmatter}

\title{Magnetism and unconventional superconductivity in Ce$_n$M$_m$In$_{3n+2m}$ heavy-fermion crystals}

\author[address1]{J. D. Thompson
\thanksref{thank1}}, \author[address1]{M. Nicklas}, \author[address1]{A. Bianchi}, \author[address1]{R.
Movshovich}, \author[address1]{A. Llobet}, \author[address1]{W.
Bao}, \author[address1]{A. Malinowski}, \author[address1]{M. F.
Hundley}, \author[address1]{N. O. Moreno}, \author[address1]{P.
G. Pagliuso}, \author[address1]{J. L. Sarrao},
\author[address2]{S. Nakatsuji}, \author[address2]{Z. Fisk},
\author[address3]{R. Borth},
\author[address3]{E. Lengyel}, \author[address3]{N. Oeschler},
\author[address3]{G. Sparn}, \author[address3]{F. Steglich}

\address[address1]{Los Alamos National Laboratory, MS K764, Los Alamos, NM 87545 USA}

\address[address2]{NHMFL, Florida State University, Tallahassee, FL 32310 USA}

\address[address3]{Max-Planck-Institute for Chemical Physics of
Solids, Dresden, D-01187 Germany}

\thanks[thank1]{ E-mail:jdt@lanl.gov}

\begin{abstract}
We review magnetic, superconducting and non-Fermi-liquid
properties of the structurally layered heavy-fermion compounds
Ce$_n$M$_m$In$_{3n+2m}$ (M=Co, Rh, Ir). These properties suggest
d-wave superconductivity and proximity to an antiferromagetic
quantum-critical point.
\end{abstract}

%
%
\end{frontmatter}

    The report \cite{hegger01} of pressure-induced superconductivity
in the heavy-fermion compound CeRhIn$_5$, with a transition
temperature exceeding 2~K, has motivated further exploration
\cite{thompson01,petrovicCo,petrovicIr,nicklas02} of this
compound and the broader family of materials
Ce$_n$M$_m$In$_{3n+2m}$, where M is a transition metal Co, Rh or
Ir . Diffraction studies \cite{grin79,mosh01} show that the
family can be considered a structural hybrid of CeIn$_3$ and
'MIn$_2$'; for $n=1$, single layers of CeIn$_3$ and 'MIn$_2$' are
stacked sequentually along the tetragonal c-axis, and for $n=2$
there are two adjacent layers of CeIn$_3$ separated by a single
layer of 'MIn$_2$'. Crystallographic layering in the $n=1$ members
produces electronic anisotropy, reflected particularly in a Fermi
surface dominated by a slightly warped cylindrical sheet.
\cite{hall01,haga01,settai01,shishido} Though these materials
should not be considered strictly 2-dimensional, their electronic
and structural anisotropies do influence magnetic,
superconducting and quantum-critical properties. In the
following, we briefly review what has been learned about some of
these properties.

\begin{table}[b]
\centering
\begin{tabular}{|c|c|c|c|c|}
  \hline
    & $T_{N}$ & Q & $\mu_{o}$ & $P_{c}$ \\
    & (K) & (h,k,l) & ($\mu_{B}$) & (GPa) \\ \hline
  CeIn$_{3}$ & 10.2 & $(\frac{1}{2},\frac{1}{2},\frac{1}{2})$ \cite{jon,benoit} & 0.65 \cite{jon} & $2.6\pm0.1$ \cite{walker97} \\
   &  &  & 0.48 \cite{benoit} &  \\ \hline
  CeRhIn$_{5}$ & 3.8 \cite{hegger01} & $(\frac{1}{2},\frac{1}{2},0.297)$ \cite{bao01a} & 0.37 \cite{bao01a} & $1.6\pm0.1$ \cite{hegger01}  \\ \hline
  Ce$_{2}$RhIn$_{8}$ & 2.8 \cite{thompson01} & $(\frac{1}{2},\frac{1}{2},0)$ \cite{bao01b} & 0.55 \cite{bao01b} & 3 $\pm$ 0.5 \cite{nicklas02} \\
   & 1.65 \cite{artur} &  &  & 0.04 $\pm$ 0.01 \cite{nicklas02} \\
  \hline
\end{tabular}
\caption{Magnetic properties of Ce$_{m}$Rh$_{n}$In$_{3m+2n}$ and
CeIn$_{3}$. $T_{N}$: N\'{e}el temperature; $Q$: antiferromagnetic
propagation vector; $\mu_{o}$: ordered moment; $P_{c}$: critical
pressure to suppress $T_{N}$ to $T=0$.}
\end{table}

\begin{table*}[t]
\centering
\begin{tabular}{|c|c|c|c|c|c|c|c|c|c|}
  \hline
    & $T_{c}$ & $ \Delta C/\gamma_n T_{c}$ & $\gamma _{n}$ & $H_{c2}^{a,b}$ & $H_{c2}^{c}$ & ${\partial H_{^{c2} }^{a,b} }/{{\partial T}}$ & ${\partial {\rm{H}}_{c2}^c }/{{\partial T}}$ & ${\rm{\xi }}_0^{a,b}$ & ${\rm{\xi }}_0^{c}$  \\
   & (K) &  & (J/mol K$^{2}$) & (T) & (T) & (T/K) & (T/K) & ($\AA$) & ($\AA$) \\
  \hline
  CeIrIn$_{5}$ & 0.4 \cite{petrovicIr} & 0.76 \cite{petrovicIr} & 0.70 \cite{petrovicIr} & 1.0 \cite{shishido} & 0.49 \cite{shishido} & -4.8 \cite{petrovicIr} & -2.54 \cite{movsh01} & 260 \cite{shishido} & 180 \cite{shishido}
  \\ \hline
  CeCoIn$_{5}$ & 2.3 \cite{petrovicCo} & 4.5 \cite{petrovicCo} & 0.35 \cite{petrovicCo} & 11.9 \cite{murphy02} & 4.95 \cite{murphy02,ikeda01} & -24.0 \cite{ikeda01} & -8.2 \cite{movsh01} & 82 \cite{ikeda01} & 53 \cite{ikeda01}
  \\
  & & & & 11.6 \cite{ikeda01} & & & -11.0 \cite{ikeda01} & &  \\  \hline
  CeRhIn$_{5}$ & 2.12 \cite{fisher01} & 0.36 \cite{fisher01} & 0.38 \cite{fisher01} &  & 10.2 \cite{muramatsu01} &  & -15.0 \cite{muramatsu01} & 57 \cite{muramatsu01} &  \\
  $@$ 2.1 GPa & & & & & at 2.5~GPa& & at 2.5~GPa& &  \\  \hline
  Ce$_{2}$RhIn$_{8}$ & 1.1 \cite{nicklas02} &  & $\approx 0.20$ \cite{nicklas02} & 5.4 \cite{nicklas02} &  & -9.2 \cite{nicklas02} &  &  & 77 \cite{nicklas02} \\
  $@$ 1.63 GPa & & & & & & & & &  \\
  \hline
\end{tabular}
\caption{Superconducting properties of
Ce$_{m}$T$_{n}$In$_{3m+2n}$. $T_{c}$: superconducting transition
temperature; $ \Delta C/\gamma_{n} T_{c}$: jump in specific heat
at $T_{c}$ normalized by the Sommerfeld coefficient $\gamma_{n}$
at $T \geq T_{c}$; $H_{c2}^{a,b}$ ($H_{c2}^{c}$): upper critical
field in the $\textit{a-b}$ plane (parallel to the c-axis)
extrapolated to $T=0$; $\partial H_{c2}/\partial T$: slope of the
upper critical field near $T_{c}$; $\xi_0$: Ginzburg-Landau
superconducting coherence length at $T=0$.}
\end{table*}

The infinite-layer (cubic), parent of this family, CeIn$_3$,
orders antiferromagnetically near 10~K at atmospheric pressure.
Applying pressure suppresses its N\'{e}el temperature toward
$T=0$ at a critical pressure $P_c\approx2.6$~GPa, where a 'dome'
of superconductivity appears in a narrow pressure window centered
around $P_c$. \cite{walker97} The single and bilayer members with
M=Rh also order antiferromagnetically and become pressure-induced
superconductors, but both with nearly an order of magnitude
higher $T_c$ \cite{hegger01,nicklas02} than the maximum of
$\sim0.25$~K found in CeIn$_3$. Some magnetic properties of these
three compounds are summarized in table 1. The commensurate
ordering $Q$-vector, ordered moment and $P_c$ are similar in
CeIn$_3$ and Ce$_2$RhIn$_8$; however, at 1.65~K, Ce$_2$RhIn$_8$
also develops an incommensurate magnetic structure \cite {artur},
as does CeRhIn$_5$. From this comparison, the $n=2$ member
superficially appears to be a magnetic hybrid of the $n=1$ and
$n=\infty$ members, as might be expected from its crystal
structure. Inelastic neutron scattering studies \cite{bao02} of
CeRhIn$_5$ find that magnetic correlations develop on a
temperature scale roughly twice $T_N$. The correlation length
along the tetragonal c-axis $\xi_c\approx1.3~{\bf c}$; whereas, in
the {\bf{a-b}} plane, the correlation length $\xi_a\approx5~{\bf
a}$, reflecting magnetic anisotropy that may be important for
superconductivity. Presently, we do not know if this anisotropy
changes as the antiferromagnetic-superconducting boundary is
approached with applied pressure, but it appears \cite{llobet}
that the c-axis discommensuration $\delta$ is somewhat pressure
dependent.

A rather remarkable characteristic of this family of materials is
their instability to superconductivity. Besides CeIn$_3$,
CeRhIn$_5$ and Ce$_2$RhIn$_8$ under pressure, CeIrIn$_5$ and
CeCoIn$_5$ are superconducting at atmospheric pressure. See table
2. In each case, superconductivity develops out of a highly
correlated state with a large specific heat Sommerfeld
coefficient $\gamma_n$ and in proximity to antiferromagnetism.
For example, substituting a small amount of Rh into
CeCo$_{1-x}$Rh$_x$In$_5$ or CeIr$_{1-x}$Rh$_x$In$_5$ induces
antiferromagnetism, and, for a range of $x$ (roughly $0.3\lesssim
x\lesssim0.7$), homogeneous coexistence of superconductivity and
antiferromagnetism. \cite{pagliuso01,Zapf,zheng02} The
superconducting transition temperatures also are high compared to
other heavy-fermion examples. With applied pressure, all compounds
listed in table 2 have $T_cs$ between 2.2 and 2.6~K
\cite{nicklas02,muramatsu01,sidorov}, except CeIrIn$_5$ whose
bulk $T_c$ reaches $\approx1$~K at 2.1~GPa \cite{nicklasun,borth}
and does not exceed 1.2~K at pressures to 4~GPa \cite{sidorovun}.

\begin{table}[b]
\centering
\begin{tabular}{|c|c|c|c|c|}
  \hline
   & $C/T$ & $\kappa $ & $1/T_{1}$ & $\lambda$ \\
  \hline
  CeIrIn$_{5}$ & $T$ \cite{movsh01} & $T^{3}$ \cite{movsh01} & $T^{3}$ \cite{zheng01,kohori01} & $T^{1.5 \pm 0.2}$
  \cite{higimoto01}
  \\ \hline
  CeCoIn$_{5}$ & $T$ \cite{movsh01} & $T^{3.37}$ \cite{movsh01} & $T^{3+\varepsilon}$ \cite{kohori01} & $T^{1.65 \pm 0.2}$ \cite{higimoto01}  \\
   &  &  &  & $T^{1+\varepsilon}$ \cite{ormento02} \\
   &  &  &  & $T^{1.5}$ \cite{ozcan} \\
   &  &  &  & $T^{\frac{3}{2}}$/$T$ \cite{chia} \\ \hline
  CeRhIn$_{5}$ & $T$ \cite{fisher01} &  & $T^{3}$ \cite{mito} &  \\
  $@$ 2.1 GPa &  &  &  &  \\
  \hline
\end{tabular}
\caption{Power laws in the superconducting state. $C/T$: specific
heat divided by temperature; $\kappa $: electronic thermal
conductivity; $1/T_{1}$: spin-lattice relaxation rate; $\lambda$:
superconducting penetration depth.}

\end{table}

Electronic anisotropy is reflected in an upper critical magnetic
field that is typically two times larger for $H\parallel
{\rm\bf{a-b}}$ plane than for $H\parallel$ c-axis. In many cases,
the measured $H_{c2}(0)$ exceeds the Pauli paramagnetic limit
$H_P/T_c=1.86$~T/K. \cite{clogston} In this regard, CeCoIn$_5$ has
been studied most extensively and, for $H\parallel [001]$,
exhibits a first order phase transition in a narrow field range
at low temperatures \cite {bianchi}, which is attributed to Pauli
limiting. For $H\parallel [110]$, a magnetically hysteretic
transition develops below $0.6T_c$ that is consistent with the
formation of a spatially inhomogeneous
Fulde-Ferrel-Larkin-Ovchinnikov state. \cite{murphy02} This
possibility deserves further study. Additionally, $H_{c2}(0)$ is
weakly, but clearly, anisotropic within the basal plane
\cite{murphy02,izawa01}, implying the possibility of non-s-wave
pairing.

There is growing evidence, summarized in table 3, that
superconductivity in members of the family is unconventional.
Power laws found deep in the superconducting state, $C/T\propto
T$, thermal conductivity $\kappa\propto T^3$ and spin-lattice
relaxation rate $1/T_1\propto T^3$, are those expected of an
order parameter with line nodes. Together with Knight shift
measurements \cite{kohori01,curro} on CeCoIn$_5$, these power
laws suggest unconventional spin-singlet superconductivity, and,
indeed, thermal conductivity measurements \cite{izawa01} find a
prominent four-fold modulation in $\kappa$ as a magnetic field is
rotated in the basal plane. The magnitude and location of maxima
in $\kappa(\theta)$ are consistent with an order parameter having
$d_{x^2-y^2}$ symmetry.

The boson mediating Cooper pairing remains unknown, but the
preponderance of evidence points to antiferromagnetic spin
fluctuations. The temperature dependence of some normal state
properties further suggests that these fluctuations may not be
conventional. For a Landau Fermi liquid, $C/T\sim {\rm
constant}$, $\rho\sim T^2$, and $1/T_1T\sim {\rm constant}$ are
expected at low temperatures. As shown in table 4, this is not
the case for several family members. These distinctly
non-Fermi-liquid behaviors are expected \cite{stewart} near an
antiferromagnetic quantum-critical point and are found for the
examples in table 4 only in $T-P-H$ space close to
superconductivity. The functional dependencies, particularly
$\rho\propto T$, suggest 2-dimensional antiferromagnetic quantum
fluctuations. Understanding the interplay of electronic and
magnetic anisotropies with quantum-critical fluctuations and
superconductivity is one problem posed by this interesting family
of heavy-fermion compounds.

\begin{table}[t]
\centering
\begin{tabular}{|c|c|c|c|}
  \hline
   & $C/T$ & $\rho $ & 1/T$_{1}$T \\
  \hline
  CeIn$_{3}$ & & $T^{1.5-1.6}$ \cite{walker97} & const. \cite{kawasaki} \\
    &  & near $P_{c}$ & $P\geq P_{c}$ \\ \hline
  CeIrIn$_{5}$ & $\gamma_0-AT^{\frac{1}{2}}$ \cite{kim} & $T^{1.3}$ \cite{petrovicIr} & $(T+8)^{-\frac{3}{4}}$
  \cite{zheng01}
  \\
    &for $H=6$~T  &  & $(T+0.86)^{-\frac{1}{2}}$ \cite{kohori01} \\ \hline
  CeCoIn$_{5}$ & $-\ln T$ \cite{petrovicCo,nakatsuji} & $T^{1.0 \pm 0.1}$ \cite{sidorov,nakatsuji} & $\sim T^{-\frac{3}{4}}$
  \cite{kohori01}
  \\ \hline
  CeRhIn$_{5}$ &  & $T ^{1}$ \cite{muramatsu01} & $(T+1.5)^{-\frac{1}{2}}$ \cite{mito}  \\
  &  &  $P = 3.2$ GPa  &   $P = 2.1$ GPa \\
  \hline
  Ce$_{2}$RhIn$_{8}$ &  & $T^{0.95 \pm 0.05}$ \cite{nicklas02} &  \\
  &  &  $P = 1.63$ GPa  &   \\  \hline
\end{tabular}
\caption{Non-Fermi-liquid behaviors. $C/T$: specific heat divided
by temperature; $\rho $: electrical resistivity in the $a-b$
plane; $1/T_{1}T$: spin-lattice relaxation rate divided by
temperature.}
\end{table}

Finally, we note the possible existence of a spin pseudogap in
CeRhIn$_5$ near its critical pressure $P_c$ \cite{kawasaki} and in
CeCoIn$_5$ for $0\leqslant P\lesssim 1.6$ ~GPa \cite{sidorov}. The
small difference in temperature scale ($\sim5$~K and $\sim3$~K,
respectively) on which a pseudogap signature appears in these two
compounds seems to be related to their relative cell volumes.
\cite{sidorov} An analysis of systematic changes in thermodynamic
and transport properties of Ce$_{1-x}$La$_x$CoIn$_5$ further
suggests a connection between the possible pseudogap in
CeCoIn$_5$ and the development of short-range antiferromagnetic
correlations. \cite{nakatsuji}
%
%


%
%
\begin{ack}
Work at Los Alamos was performed under the auspices of the U.S.
DOE Office of Science. ZF acknowledges support by NSF grant
DMR-9971348. We also thank V. A. Sidorov and H. A. Borges for
communicating results of their unpublished pressure measurements.
\end{ack}

%
%

\end{document}